\documentclass[prd,showpacs,preprintnumbers,floatfix,superscriptaddress,10pt]{revtex4}
\usepackage[mathscr]{eucal}
\usepackage[dvips]{color}
\usepackage[dvips]{graphicx}
\usepackage{epsf}
\usepackage{bm}
\usepackage{amssymb}
\usepackage{amsmath}
\usepackage[normalem]{ulem}
\newcommand{\be}{\begin{equation}}
\newcommand{\ee}{\end{equation}}
\newcommand{\ba}{\begin{eqnarray}}
\newcommand{\ea}{\end{eqnarray}}
\newcommand{\ban}{\begin{eqnarray*}}
\newcommand{\ean}{\end{eqnarray*}}

\newcommand{\ie}{{\it i.e.\,}}

\newcommand{\bef}{\begin{figure}}
\newcommand{\eef}{\end{figure}}
\newcommand{\bce}{\begin{center}}
\newcommand{\ece}{\end{center}}

\begin{document}

\title{Jet-induced gauge field instabilities in the quark-gluon plasma: A kinetic theory approach }
\author{Massimo~Mannarelli}
\author{Cristina~Manuel}
\affiliation{Instituto de Ciencias del Espacio (IEEC/CSIC),
Campus Universitat Aut\`onoma de Barcelona, Facultat de Ci\`encies, Torre C5 E-08193 Bellaterra (Barcelona), Spain}
\date{\today}
\begin{abstract}
We discuss the  properties of the collective modes of a system composed by a thermalized quark-gluon plasma traversed by a relativistic jet of partons. The  transport equations obeyed by the components of the plasma and of the  jet are studied  in the Vlasov approximation. Assuming  that the  partons in the jet  can be described with a tsunami-like distribution function we derive the expressions 
of the dispersion law of the collective modes. Then the behavior of the unstable gauge modes of the system is analyzed for various values of the velocity of the jet, of the momentum of the collective modes and of the angle between these two quantities. We find that  the most unstable modes are those with momentum orthogonal to the velocity of the jet and that these instabilities  appear when the velocity of the jet is higher
than a threshold value, which depends on the plasma and jet frequencies. The results obtained within the Vlasov approximation are compared with the corresponding results obtained   using a chromohydrodynamical approach.
The effect we discuss here suggests a possible collective mechanism for the description of the jet quenching phenomena in heavy ion collisions. 
\end{abstract}
\preprint{}
 \maketitle

\section{Introduction}

One of the interesting aspects of ultrarelativistic heavy-ion
collisions is that the properties of the produced  matter can be studied employing   high $p_T$ partons generated  by hard scatterings in the initial stage of the collision. 
Such   high $p_T$ partons   behave as  hard probes unveiling some properties of the system and providing  evidences of the production  of
a thermalized quark-gluon plasma (QGP) \cite{Adams:2005dq}.
When the jet of partons travels across the medium it loses  energy  and degrades, mainly by radiative processes (see \cite{Kovner:2003zj} for reviews). The energy and momentum of the jet are   absorbed   by the plasma and result in an increased  production of  soft hadrons in the direction of propagation of the partons. 
The description of this process is currently undertaken employing  various models,  
 where perturbative QCD is supplemented with medium-induced parton energy loss \cite{Eskola:2004cr} or where the AdS/CFT correspondence \cite{Liu:2006ug} is employed.

In a previous paper \cite{Mannarelli:2007gi} we have proposed a novel mechanism for describing how the jet loses energy and momentum while traveling in a thermally equilibrated   quark-gluon plasma.  
Since the jet of particles  is  not in thermal equilibrium with the QGP it perturbs and destabilizes the system  inducing the generation of gauge fields.  Some of these gauge modes are unstable and grow  exponentially fast in time absorbing the kinetic energy of the  jet. A basic assumption in \cite{Mannarelli:2007gi}   was that that  both the plasma and the jet  can be described using a fluid approach. This  approach developed in \cite{Manuel:2006hg} has been derived from kinetic theory expanding the transport equations in moments of momenta and truncating the expansion at the second moment level. The system of equations is then closed with an equation of state relating  pressure and energy density. 
The fluid  approach  has several advantages with respect to the underlying kinetic theory. The most remarkable one is that  one has to deal with  a set of  equations much simpler than those of kinetic theory. Then one can    easily generalize the fluid equations  to  deal with more complicated systems. This is a strategy that has been successfully followed in the  study of
different dynamical aspects of non-relativistic electromagnetic plasmas \cite{Kra73}.

In the present paper we consider  the same setting of our previous paper \cite{Mannarelli:2007gi}, \ie  a static plasma traversed by a relativistic jet, but we use kinetic theory  instead of the fluid approach. The appearance of filamentation instabilities due
to hard jets was first pointed out in Ref. \cite{Pavlenko:1991ih}. Here we  generalize that study, 
also with the aim of preparing the ground for the computation of  the jet energy loss. However, in the present paper we content ourselves with the study of the collective modes of the system,   postponing to a future publication \cite{future}  the computation  of the  jet energy loss.

Transport theory provides a well controlled framework for studying the properties of the quark-gluon plasma
in the weak coupling regime, $g \ll 1$.
Indeed it is well known that the physics of long distance scales in an equilibrated weakly coupled
QGP can be described within semiclassical transport equations \cite{Braaten:1989mz,Bla93,Kel94}. In this approach the
hard modes,  with typical energy scales of order $T$,  are treated as (quasi-)particles which propagate in the background of the soft modes, whose energies are equal or less than $gT$,
which are treated as classical gauge fields.
This program has been very successful for  understanding
 some dynamical aspects of the soft gauge fields in an almost equilibrated QGP \cite{Blaizot:2001nr,Litim:2001db}.
The same approach has been recently used to study the development of gauge field
instabilities when the distribution function of the hard modes is far away from
equilibrium, and it is anisotropic in momenta space (see Ref.~\cite{Mrowczynski:2006ad} and references therein). We will also use the
semiclassical transport approach in order to study how the soft gauge field dynamics
of a thermally equilibrated QGP is perturbed by a relativistic jet
of particles.   

The hard modes of the QGP can be described with two different transport equations. One can treat the color charges as classical variables \cite{Hei83,Litim:2001db} or  one can treat color
as a quantum degree of freedom \cite{Elze:1989un,Mrowczynski:np}. 
We will employ the latter formulation. However since we will study the gauge fields dynamics in the linear response approximation employing the former formulation we would have obtained the same results. Indeed in the leading response analysis the two possibilities lead to the same answers. 
 We will also compare the results for the growing rates of the unstable modes  obtained in transport theory with the analogous results we obtained with the fluid approach in Ref.  \cite{Mannarelli:2007gi}.

This paper is organized as follows. In Section \ref{vlasov} we review the kinetic theory approach in the so-called Vlasov approximation and specialize the system of equations to the case of interest of a plasma traversed by a jet of particles. In Section \ref{QGP} we consider separately a thermally equilibrated plasma and a jet of particles with a so-called tsunami-like distribution function.  In both cases we derive the dispersion laws of the collective modes and find that the two systems are stable.
In Section \ref{QGP-JET} we show that in a system composed by a thermalized plasma traversed by a jet of
particles unstable collective modes appear, and we determine their growing rate.
 We discuss separately the cases where the momentum of the collective mode is collinear with the  velocity of the jet, orthogonal or at an arbitrary angle.  
In Section \ref{comparison} the results obtained within the kinetic theory approach are   compared with the results  of  the fluid approach. 
In Section \ref{conclusion} we discuss our results and comment on how our study could be improved to make contact with heavy ion phenomenology.

\section{Kinetic theory for colored particles in the Vlasov approximation}
\label{vlasov}

In this Section we review  the  transport equations obeyed by the
distribution functions of  colored particles \cite{Elze:1989un,Mrowczynski:np} that  represent  the starting point of our analysis.

The distribution functions of quantum colored particles
are  hermitian matrices, whose dimensionality depends on
the  color representation  of the particle. They are not gauge invariant quantities, but rather transform
covariantly under a gauge transformation.
Quarks and antiquarks belong to the fundamental representation of $SU(3)_c$ and the corresponding distribution functions $Q(p,x)$ and $\bar Q(p,x)$ are $3 \times 3$ matrices in color space. Gluons are in the adjoint representation of $SU(3)_c$ and  their distribution function $G(p,x)$ is a $8 \times 8$ matrix in color space. Without loss of generality we will consider
that the particles in the jet are quarks, and we will denote the corresponding distribution function as
$W_{\rm jet} (p,x)$, separating explicitly that component from $Q(p,x)$. The reason for doing so is that we
will consider very different initial conditions for the thermally equilibrated quarks in the plasma and for the quarks in the jet. Also, one should note that the collisional rates will probably be very different for these two systems of quasiparticles. 

The distribution function of quarks, antiquarks, gluons and of the particles of the jet satisfy the following transport equations 
\ba
p^{\mu} D_{\mu}Q(p,x) + {g \over 2}\: p^{\mu}
\left\{ F_{\mu \nu}(x), \partial^\nu_p Q(p,x) \right\}
&=& C \;,
\label{transport-q}  \\ [2mm]
p^{\mu} D_{\mu}\bar Q(p,x) - {g \over 2} \: p^{\mu}
\left\{ F_{\mu \nu}(x), \partial^\nu_p \bar Q(p,x)\right\}
&=& \bar C \;,
\label{transport-barq} \\ [2mm]
p^{\mu} {\cal D}_{\mu}G(p,x) + {g \over 2} \: p^{\mu}
\left\{ {\cal F}_{\mu \nu}(x), \partial^\nu_p G(p,x) \right\}
&=& C_g \;,
\label{transport-gluon}
\\
p^{\mu} D_{\mu}W_{\rm jet}(p,x) + {g \over 2}\: p^{\mu}
\left\{ F_{\mu \nu}(x), \partial^\nu_p W_{\rm jet} (p,x) \right\}
&=& C_W\,.
\label{transport-jet}
\ea
With  $\{...,...\}$ we denote the anticommutator, $\partial^\nu_p$ is
the four-momentum derivative and $g$ is the QCD coupling constant.  The covariant derivatives $D_{\mu}$ and ${\cal D}_{\mu}$ act as
$$
D_{\mu} = \partial_{\mu} - ig[A_{\mu}(x),...\; ]\;,\;\;\;\;\;\;\;
{\cal D}_{\mu} = \partial_{\mu} - ig[{\cal A}_{\mu}(x),...\;]\;,
$$
with $A_{\mu }=  A^{\mu }_a (x) \tau^a$ and ${\cal A}_{\mu }=  A^{\mu }_a (x) T^a$, and $\tau^a$ and $T^a$ are $SU(3)$ generators in the fundamental and adjoint representations, respectively.
The strength tensor in the fundamental representation is
$F_{\mu\nu}=\partial_{\mu}A_{\nu} - \partial_{\nu}A_{\mu}
-ig [A_{\mu},A_{\nu}]$, while  ${\cal F}_{\mu \nu}$ denotes the field
strength tensor in the adjoint representation. 

In Eqs. (\ref{transport-q}),(\ref{transport-barq}), (\ref{transport-gluon}) and
(\ref{transport-jet})
 $C, \bar C, C_g$ and $C_W$ represent the collision terms. For time scales shorter than the mean free path time the collision terms  can be neglected, as typically done in the so-called Vlasov approximation. 
The knowledge of the distribution function allows one to compute the associated color current, which in a self-consistent
treatment enters as a source term in the Yang-Mills equation. 
Therefore in this approximation  the different components of the system formed by the plasma and the jet interact with each other only through the generated average gauge fields.

Assuming that the quark-gluon plasma is initially in a  colorless and thermally equilibrated  state, we can describe the small deviation  from  equilibrium of the distribution functions of  quarks, antiquarks and gluons   as
\be
Q(p,x) = f^{\rm eq.}_{FD}(p_0) + \delta Q(p,x) \ , \qquad \bar Q(p,x) = f^{\rm eq.}_{FD}(p_0) + \delta \bar Q(p,x) \ ,
\qquad
G(p,x) = f^{\rm eq.}_{BE}(p_0) + \delta G(p,x) \ ,
\ee
where
\be
f^{\rm eq.}_{FD/BE}(p_0) = \frac{1}{e^{p_0/T} \pm 1}
\ee
are the Fermi-Dirac and Bose-Einstein equilibrium distribution functions.

We will also study color fluctuations of
the initial colorless distribution function of the jet. Thus
\be
W_{\rm jet}(p,x) = f_{\rm jet} (p) + \delta W_{\rm jet}(p,x) \,.
\ee
For the initial jet distribution function we will consider a colorless tsunami-like form \cite{Pisarski:1997cp}
\be
\label{tsunami}
f_{\rm jet}(p) =  \bar n \, \bar u^0 \;
\delta^{(3)}\Big({\bf p}
- \Lambda \, \bar {\bf u} \Big) \;,
\ee
that describes a system of particles of   constant density $\bar n$, all moving with the same  velocity $\bar u^\mu = (\bar u^0, \bar {\bf u}) = \gamma (1, {\bf v})$, where $\gamma$ is the Lorentz factor.    The parameter $\Lambda$ fixes the scale of the energy of the particles. 

Although this distribution function is adequate for describing a uniform and sufficiently dilute system of particles, it would be very interesting to extend our analysis to more complicated  forms. Using more involved distribution functions   on the one hand  would lead to a more accurate description of the jets  relevant for heavy ion phenomenology, 
when the density of particles composing the jet is not uniform and there is a spread in momentum. However, on the other hand this would  complicate the dispersion laws of the collective modes. For this reason   we  leave such analysis to a future work.

In the Vlasov approximation one can compute the contribution to the  polarization tensor
of  the particles of species $\alpha$ (where $\alpha$ refers to quarks, antiquarks, gluons or
the partons of the jet)
 as
\be
\label{Pi-kinetic}
\Pi^{\mu \nu}_{ab, \alpha}(k) = - g^2 C^{\alpha}_F \delta_{ab}\int_p
f_{\alpha} (p) \;
{ (p\cdot k)(k^\mu p^\nu + k^\nu p^\mu) - k^2 p^{\mu} p^{\nu}
- (p\cdot k)^2 g^{\mu\nu} \over(p\cdot k)^2} \;,
\ee
where $a,b$ are color indices and $C^\alpha_F$ is the value of the quadratic Casimir associated with the particle specie $\alpha$ which takes values
$1/2$ and $3$ for the fundamental and adjoint representations, respectively.
The momenta measure is defined as
\be
\label{measure}
\int_p \cdots \equiv \int \frac{d^4 p}{(2\pi )^3} \:
2 \Theta(p_0) \delta (p^2-m^2_\alpha) \;,
\ee
where $m_\alpha$ is the mass of the particle of specie $\alpha$.
For simplicity we  assume that the particles belonging to the plasma are massless.  Instead the particles of  the jet  have a non-vanishing mass.

When $f_\alpha(p)$ is a thermal equilibrated distribution function, Eq.~(\ref{Pi-kinetic})
reduces to the form of the hard thermal loop (HTL) polarization tensor \cite{Braaten:1989mz,Bla93,Kel94}.  However  for the tsunami-like distribution function, Eq.~(\ref{tsunami}),
the polarization tensor obviously takes a different form.

The gauge fields  obey the Yang-Mills equation 
\be
\label{yang-mills}
D_{\mu} F^{\mu \nu}(x) = \delta j^\nu_t (x) = \delta j_{p}^{\nu}(x) + \delta j_{\rm jet}^{\nu }(x)\; ,
\ee
where we have defined
\be
\label{col-current}
\delta j^{\mu }_p(x) = -\frac{g}{2} \int_p p^\mu \;
\Big[ \delta Q( p,x) - \delta \bar Q ( p,x)+  2 \tau^a {\rm Tr}\big[T^a \delta G(p,x) \big]\Big] \; ,
\ee
which describes the plasma color current, and
\be
\label{jetcol-current}
\delta j_{\rm jet}^{\mu }(x) = -\frac{g}{2} \int_p p^\mu \;
 \delta W_{\rm jet}( p,x)  ,
\ee
which describes the fluctuations of the  current associated with the jet.

Equation (\ref{yang-mills}) together with Eqs.~(\ref{transport-q}-\ref{transport-jet})
form a set of equations that has to be solved  self-consistently. Indeed the gauge fields which are solutions of the Yang-Mills
equation enter into the  transport equations of every particle species and,  in turn, affect the evolution of the distribution functions.

We will show that  in the presence of a fast moving jet of particles some unstable gauge modes appear, as  first pointed out in Ref.~\cite{Pavlenko:1991ih}, and also  discussed using a chromohydrodynamical approach in our previous paper  \cite{Mannarelli:2007gi}. However it is well-known that if there is no jet of particles propagating in the plasma, then  all the gauge collective modes are stable \cite{LeBellac}. This is due to the fact that the HTL approximation is obtained employing  a distribution function corresponding to a state of stable minimum for the plasma.  In the next Section we will derive the equation of the collective modes   of a thermally equilibrated QGP (where there is no jet of particles). Then we will derive the dispersion laws for the collective modes of  a jet of particles with a tsunami-like distribution function and show that also in this case the collective modes are stable.

\section{Collective modes in the equilibrated QGP and in the jet}
\label{QGP}

The collective modes of a thermally equilibrated plasma
are obtained from the knowledge of the HTL polarization tensor.
We briefly recall some results of  the HTL approach and derive the dispersion laws that describe the evolution of  gauge collective modes \cite{LeBellac}.

The dielectric tensor is defined as
\be
\label{dielectric}
 \varepsilon^{ij}(\omega,{\bf k}) = \delta^{ij} + \frac{\Pi^{ij}}{\omega^2}  \,,
\ee
where in order not to overcharge the notation we dropped the color indices.
For the QGP in the HTL approximation the dielectric tensor is given by \cite{LeBellac}
\be
\varepsilon^{ij}_{\rm p}(\omega,{\bf k}) =\Big(\delta^{ij} - \frac{k^ik^j}{k^2} \Big)\varepsilon_{\rm T} (\omega,{\bf k})+  \frac{k^ik^j}{k^2} \varepsilon_{\rm L} (\omega,{\bf k}) \ ,
\ee
where 
\ba
\varepsilon_{\rm L}(\omega,{\bf k}) &=& 1 + \frac{3\omega_{\rm p}^2}{k^2}\left[1+\frac{\omega}{2 k} \log{\frac{\omega -k}{\omega +k}}\right] \ , \\
\varepsilon_{\rm T}(\omega,{\bf k}) &=& 1 - \frac{3\omega_{\rm p}^2}{2k^2}\left[1+\left(\frac{\omega}{2 k} - \frac{k}{2 \omega}\right) \log{\frac{\omega -k}{\omega +k}}\right] \ ,
\ea
are the longitudinal and transverse (with respect to $\bf k$) components of the dielectric tensor of the plasma,
composed by gluons and massless quark  of $N_F$ flavors,
and $\omega_p^2= \frac 19 g^2T^2 (3 + N_F/2)$ is the plasma frequency squared. 

The dispersion laws  are  obtained by solving the equation 
\be
\label{general-disp}
 {\rm det}\Big[ {\bf k}^2 \delta^{ij} -k^i  k^j
- \omega^2 \varepsilon^{ij}_{\rm p}(k)  \Big]  = 0 \,,
\ee
that leads to
\ba
\label{tranHTL} k^2 - \omega^2 \varepsilon_{\rm T } &= &0 \,, \\
\label{longHTL}\varepsilon_{\rm L} &=&0 \ ,
\ea
for  the  transverse and longitudinal modes respectively. In general these equations have to be solved numerically, but simple analytical solutions can be found in two limits.
In the long wavelength limit $k \ll \omega_p$ one obtains
\ba
\label{tranHTLexp}\omega_{\rm T}^2& =& \omega_{p}^2+ \frac{6}{5}  k^2 \ , \\
\label{longHTLexp}\omega_{\rm L}^2& =& \omega_{p}^2+ \frac{3}{5}  k^2 \,,
\ea
while in the opposite situation,  $k \gg \omega_p$, one gets
\be
\label{tranHTLexp-long}\omega_{\rm T} = k  \ , \qquad  \omega_{\rm L} =   k  \ ,
\ee
which correspond to the dispersion law of free propagators. Anyway, in the  ultraviolet regime the
gauge field propagators receive  quantum corrections which are not accounted
for by the HTL or kinetic theory approximation. For those scales these two approaches are clearly not
valid. 

Now we consider the collective modes which are present in a system described by the tsunami distribution function (\ref{tsunami}) alone. These have been considered in Ref.~\cite{Pisarski:1997cp}. Upon substituting the tsunami-like distribution function Eq.~(\ref{tsunami}) in  Eq.~(\ref{Pi-kinetic}) one deduces that
\be
\Pi^{ij}_{\rm jet}(k)= -\omega_{\rm jet}^2 \left( \delta^{ij} + \frac{k^i v^j +k^j v^i}{\omega - {\bf k \cdot v} } -  \frac{(\omega^2 - k^2)v^i v^j }{(\omega - {\bf k \cdot v)^2} }  \right) \ ,
\ee
where 
the plasma frequency of the jet is given by $\omega_{\rm jet}^2 = \frac {g^2}{2} \frac{\bar n}{\Lambda}$. 

Upon substituting this expression of the polarization tensor in Eq.~(\ref{general-disp}), but using  the dielectric tensor
associated with the jet $\varepsilon_{\rm jet}^{ij}$, one obtains in a straightforward  way  the dispersion laws of the collective modes of the system. These will in general depend on $|\bf k|$, $|\bf v|$ and $\cos\theta={\bf \hat k \cdot \hat v}$. For any value of these parameters we find that the system is stable. 
One can obtain analytical expression of the collective modes for $\bf k \parallel v$, or  $\bf k \perp v$, or $v =1$. In the last case 
 we find the following non-trivial solutions 
\be
\omega^2 = \omega^2_{\rm jet} + k^2 \ , \qquad \omega = k \cos\theta \, .
\ee

\section{Collective modes in the system composed by the QGP and jet}
\label{QGP-JET}

We now consider the system composed by  an equilibrated QGP traversed by a jet of particles. We are interested in very short time scales when the Vlasov approximation can be employed. The  effect of the beam of particles is to induce a color current, which provides a contribution to
the polarization tensor. Therefore  apart from the HTL polarization tensor, we have to include the polarization tensor due to the jet. In this very short time regime, the polarization tensor of the whole system is additive, meaning that
\be
\Pi^{\mu \nu}_{t}(k) = \Pi^{\mu \nu}_{p}(k) + \Pi^{\mu \nu}_{\rm jet}(k) \ .
\ee

The total dielectric tensor is now obtained considering both the effects of the plasma and the jet
\be
\label{total-dielectric}
 \varepsilon^{ij}_{\rm t}(\omega,{\bf k}) = \delta^{ij} + \frac{\Pi^{ij}_t}{\omega^2}  \,,
\ee
and the dispersion laws of the collective modes of the whole system  can  be determined solving  the equation
\be
\label{dispersion-T}
 {\rm det}\Big[ {\bf k}^2 \delta^{ij} -k^i  k^j
- \omega^2 \varepsilon^{ij}_{\rm t}(k)  \Big]  = 0 \,.
\ee

The solutions of this equation depend on ${|\bf k|}$, ${|\bf v|}$,
$\cos\theta={\bf \hat k \cdot \hat v}$, and on
\be
 \omega_t^2 = \omega^2_{\rm p}+\omega^2_{\rm jet}
\ee 
and
\be
b = \frac{\omega^2_{\rm jet}}{\omega^2_{t}} \ .
\ee
We will analyze separately the cases where ${\bf k} \parallel {\bf v} $,  ${\bf k} \perp  {\bf v} $ and the case of arbitrary values of the angle between ${\bf k}$ and ${\bf v} $. 

As we discussed in the previous section, when the plasma and the jet do not interact, they have stable collective modes.
However, once we consider the composed system of plasma and jet interacting via mean gauge field interactions, unstable gauge modes may appear. In the following subsections we will analyze under which circumstances the system becomes unstable.

\subsection{{\bf k} parallel to {\bf v}  }\label{longsec}
Here we consider the case where the momentum of the collective mode is collinear with the velocity of the jet.  The dispersion law of the unstable mode 
can be obtained from Eq.~(\ref{dispersion-T}) upon setting $\theta = 0$
and  corresponds to one of the solutions of the equation
\be
\label{eqparallel}
  \omega_{\rm jet}^2(1-v^2)-\varepsilon_{\rm L}(\omega - kv)^2=0\,.
\ee
This equation  admits  a solution of the form $\omega = a + i \Gamma$,
corresponding to an unstable propagating mode. We find that such an instability  is present for any value of the velocity $v<1$ and for any value of the momentum $k$.  

\begin{figure}[!th]
\includegraphics[width=3.5in,angle=-0]{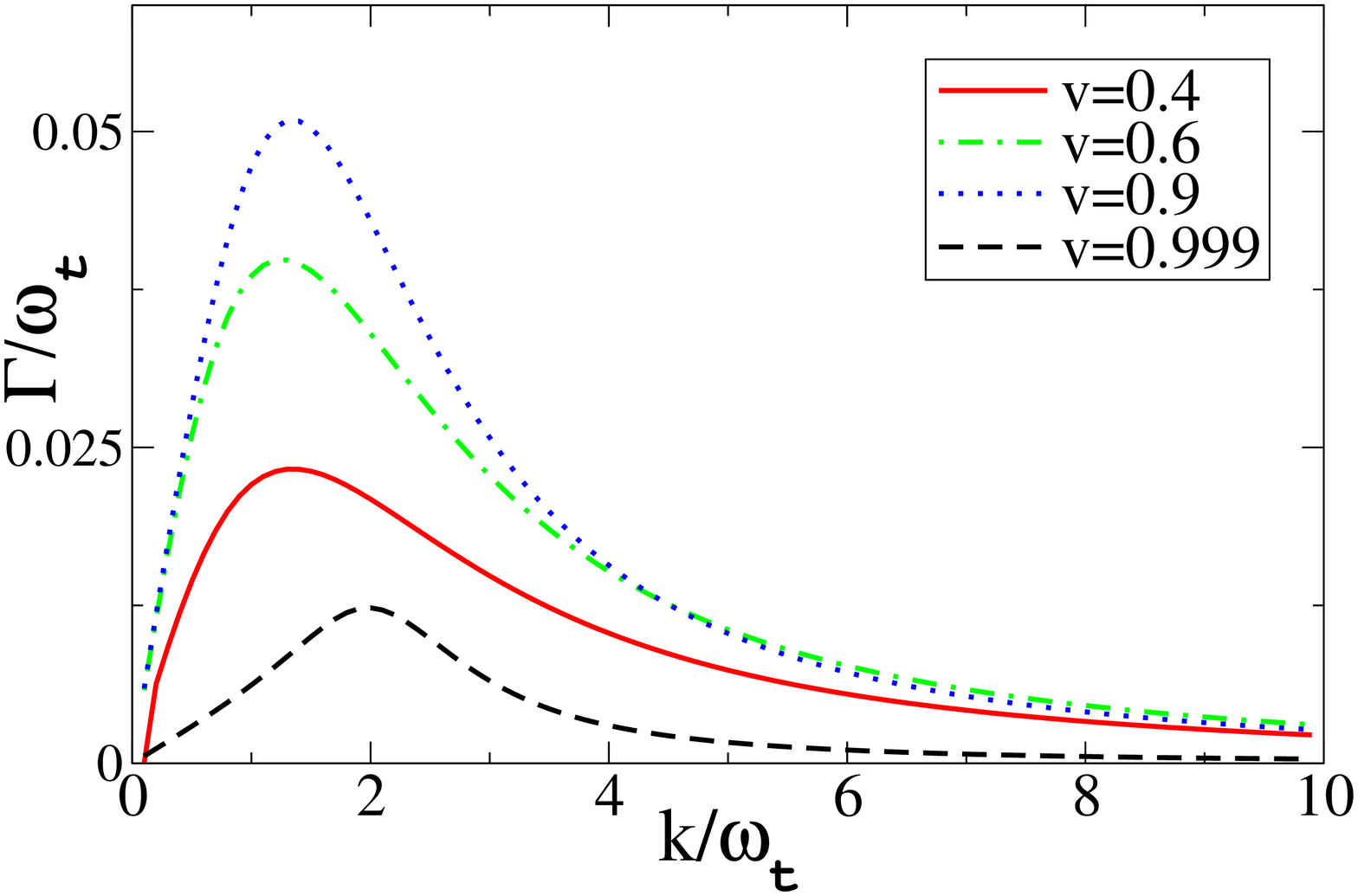}
\includegraphics[width=3.5in,angle=-0]{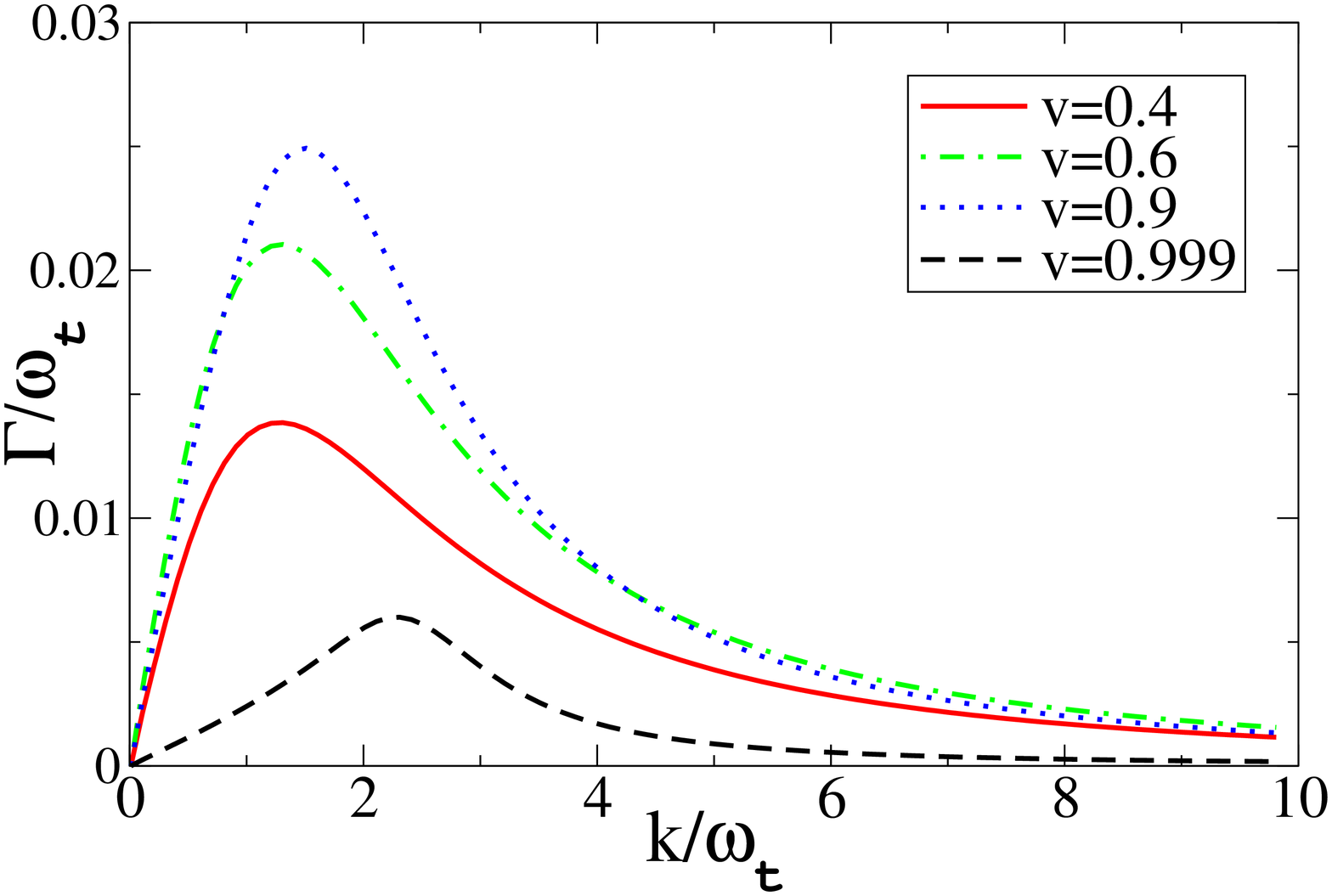}
\caption{(color online) Imaginary part of the dispersion law of the unstable longitudinal mode for the system composed by a plasma and a jet in the case $\bf k \parallel v$ as a function of the momentum of the mode at  $b=0.1$  (left) and at $b=0.02$ (right) for four different values of the velocity of the jet,  $|{\bf v}|$. \vspace{1cm}} \label{Parafig1}
\end{figure}

In Fig.~\ref{Parafig1} we present the plot of the imaginary part of the dispersion law of the unstable collective mode as a function of the momentum for four different values of the velocity of the jet and for two different values of the parameter $b$. The left panel corresponds to $b=0.1$ and the right panel corresponds to  $b=0.02$.  In both cases the largest value of $\Gamma$ is obtained  for velocities of the jet $v \sim 0.9$. The corresponding value of the momentum is $k \approx 2 \omega_t $ and is approximately independent of $b$.
In agreement with the finding of  Ref.~\cite{Mannarelli:2007gi}, in the $v \rightarrow 1$ limit, which corresponds to considering
the massless limit for the particles of the jet, this mode becomes stable. This can be easily understood from Eq.~(\ref{eqparallel}). Indeed for $v=1$, the dispersion law of the longitudinal modes are not affected by the presence of the jet, because the first term on the left hand side of  Eq.~(\ref{eqparallel}) vanishes.  Therefore in this limit the jet does not have any effect  on these modes and the system becomes stable. 

It is interesting to notice that with increasing values of $b$ the value of $\Gamma$ at the maximum does in general increase. The qualitative reason is that with increasing values of $b$ the plasma frequency of the jet increases and therefore the destabilizing effect of the jet becomes larger.
Indeed it is clear that  setting $b=0$ is equivalent to neglect the presence of the jet at all, as can be easily deduced from Eq.~(\ref{eqparallel}). With increasing $b$ (at fixed $\omega_p$),  the first term on the  left hand side of  Eq.~(\ref{eqparallel})
becomes larger and the corresponding influence on the solution of the equations grows as well. The value of the momentum corresponding to the maximum of $\Gamma$ is instead approximately independent of $b$.

\subsection{ {\bf k} orthogonal to {\bf v}}\label{orthosec}
The dispersion law of the unstable mode can be obtained  from Eq.~(\ref{dispersion-T}) upon setting $\theta = \pi/2$ and corresponds to one of the solutions
of the equation
\be
\Big[k^2-\omega^2 \varepsilon_{\rm T} + \omega_{\rm jet}^2(1-v^2+\frac{k^2v^2}{\omega^2})\Big]\Big[-\omega^2 \varepsilon_{\rm L}+\omega_{\rm jet}^2\Big] -\omega_{\rm jet}^4\frac{k^2v^2}{\omega^2}=0 \,.
\label{ortodisp}
\ee
We notice that the solutions of Eq.(\ref{ortodisp}) are non propagating, in the sense that they are pure imaginary  of the form $\omega = i \Gamma$. 
 
\begin{figure}[!th]
\includegraphics[width=3.5in,angle=-0]{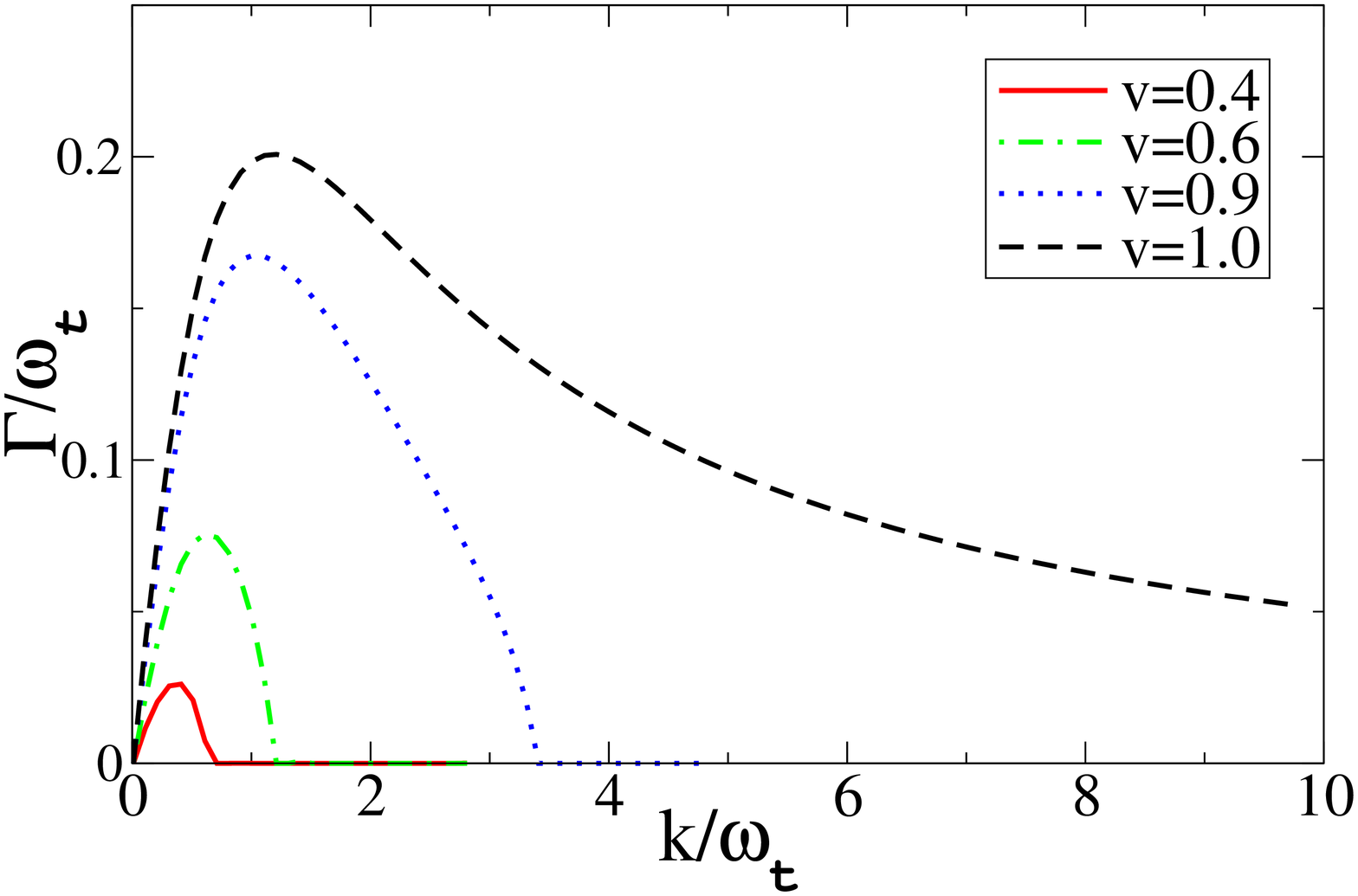}
\includegraphics[width=3.5in,angle=-0]{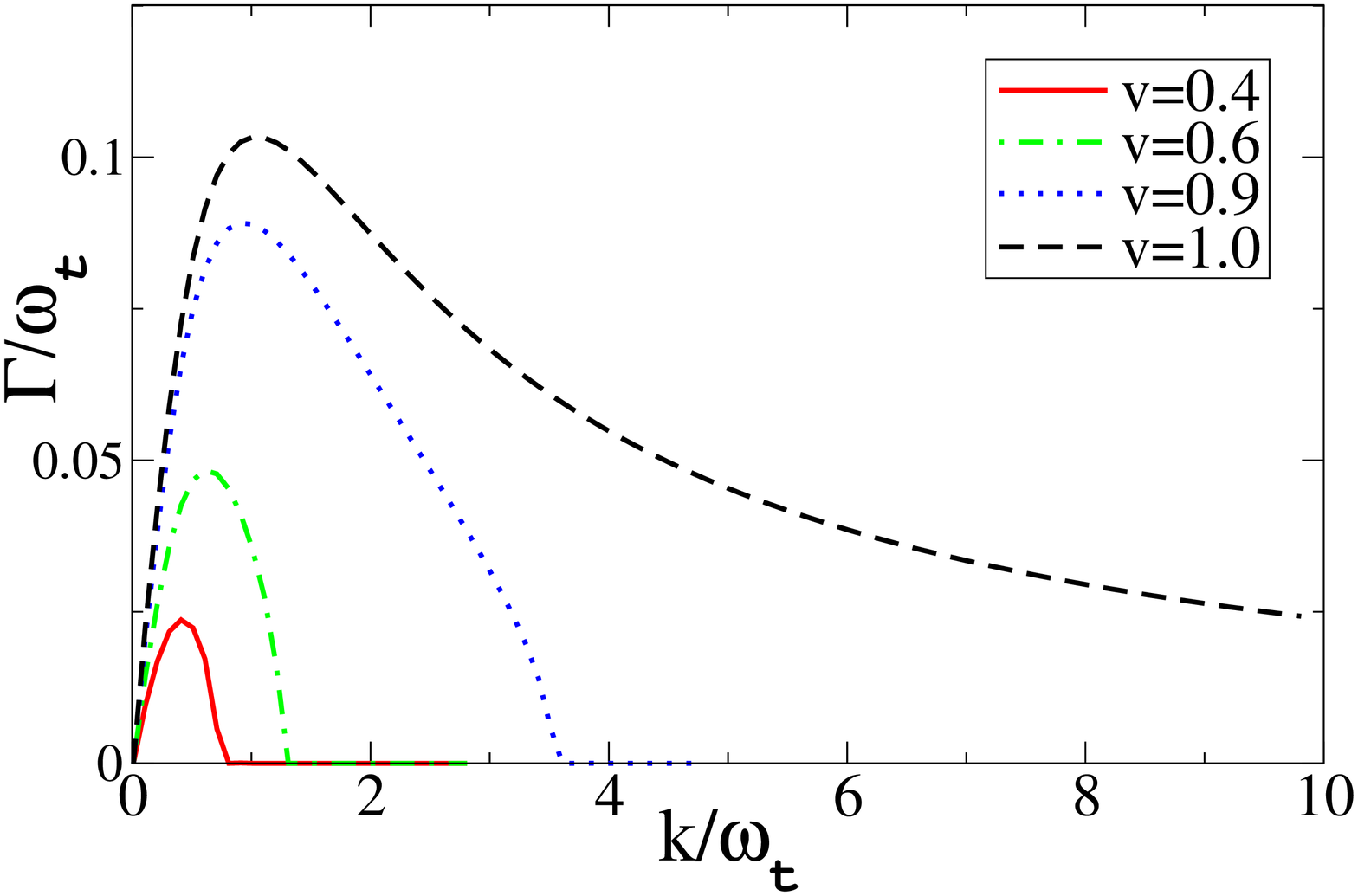}
\caption{(color online) Imaginary part of the dispersion law of the unstable  mode  for the system composed by a plasma and a jet in the case $\bf k \perp v$ as a function of the momentum of the mode at $b=0.1$  (left) and at $b=0.02$ (right) for four different values of the velocity of the jet,  $|{\bf v}|$. \vspace{1cm}}  \label{Orthofig1}
\end{figure}

We have numerically solved Eq.(\ref{ortodisp}) and reported in Fig.\ref{Orthofig1}   the plot of the imaginary part of the dispersion law of the unstable collective mode as a function of the momentum. 

As for the longitudinal case we find that with increasing values of $b$, the values of $\Gamma$ at the maximum increase and the corresponding value of the momentum remains approximately the same.
Differently from the longitudinal case, there exists a maximum value of the momentum, $k_{\rm max}$, such that for values of the momentum $k>k_{\rm max}$ there is no unstable solution. This quantity  can be obtained by solving  Eq.(\ref{ortodisp}) in the limit $\Gamma \to 0$ and is given by the expression 
\be
k_{\rm max} =  \omega_t  \sqrt{\frac{3 v^2-2 v^2 b -b }{1-v^2}} \,.
\ee
Requiring that $k_{\rm max} $ is real, we find that there is an unstable mode for
\be
b \le b_{\rm max}(v) = \frac{3 v^2}{1+2 v^2} \,,
\ee
or equivalently for 
\be\label{vmin}
v \ge v_{\rm min}(b)=\sqrt\frac{b}{3 - 2 b} \,.
\ee
Therefore there is a threshold value for  the velocity of the jet, that depends on $b$. 
If the velocity of the jet is smaller than $v_{\rm min}(b)$, then no unstable mode with momentum orthogonal  to $\bf v$ appears. Notice that the minimum value of the velocity vanishes for vanishing values of $b$.

\subsection{Arbitrary angles}\label{obliqsec}
We shall now consider various values of the angle between the velocity of the plasma and the momentum of the collective mode.  The dispersion laws are obtained solving Eq.~(\ref{dispersion-T}) for arbitrary  values of $\theta$.   Equation (\ref{dispersion-T}) can be rewritten as follows:
\be \label{eqgeneral}
(k^2-\omega^2 \varepsilon_{\rm T} + \omega_{\rm jet}^2 A^{xx})(-\omega^2 \varepsilon_{\rm L}+\omega_{\rm jet}^2A^{zz}) - \omega_{\rm jet}^4(A^{xz})^2=0
\ee
where
\ba
A^{xx} &=& 1 - \frac{(\omega^2-k^2)v^2\sin^2\theta}{(\omega-  k v \cos \theta)^2} \nonumber\,,\\
A^{xz} &=& \frac{k v \sin \theta}{\omega-  k v \cos \theta} - \frac{(\omega^2-k^2)v^2\sin \theta\cos\theta}{(\omega-  k v \cos \theta)^2} \,,\\
A^{zz} &=&  \frac{\omega^2(1-v^2\cos^2\theta)}{(\omega-  k v \cos \theta)^2} \nonumber \,.
\ea

\begin{figure}[!th]
\includegraphics[width=2.3in,angle=-0]{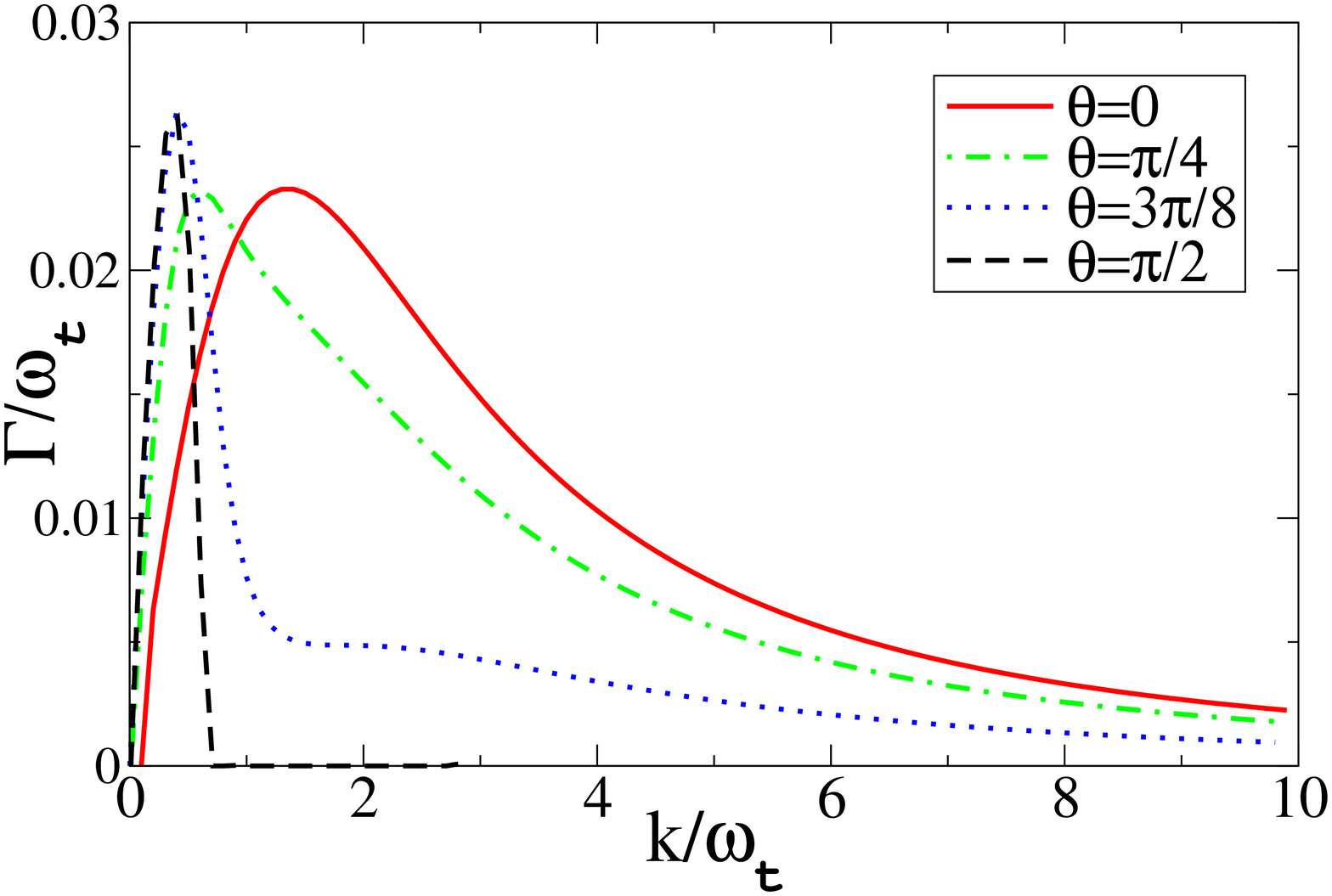}
\includegraphics[width=2.3in,angle=-0]{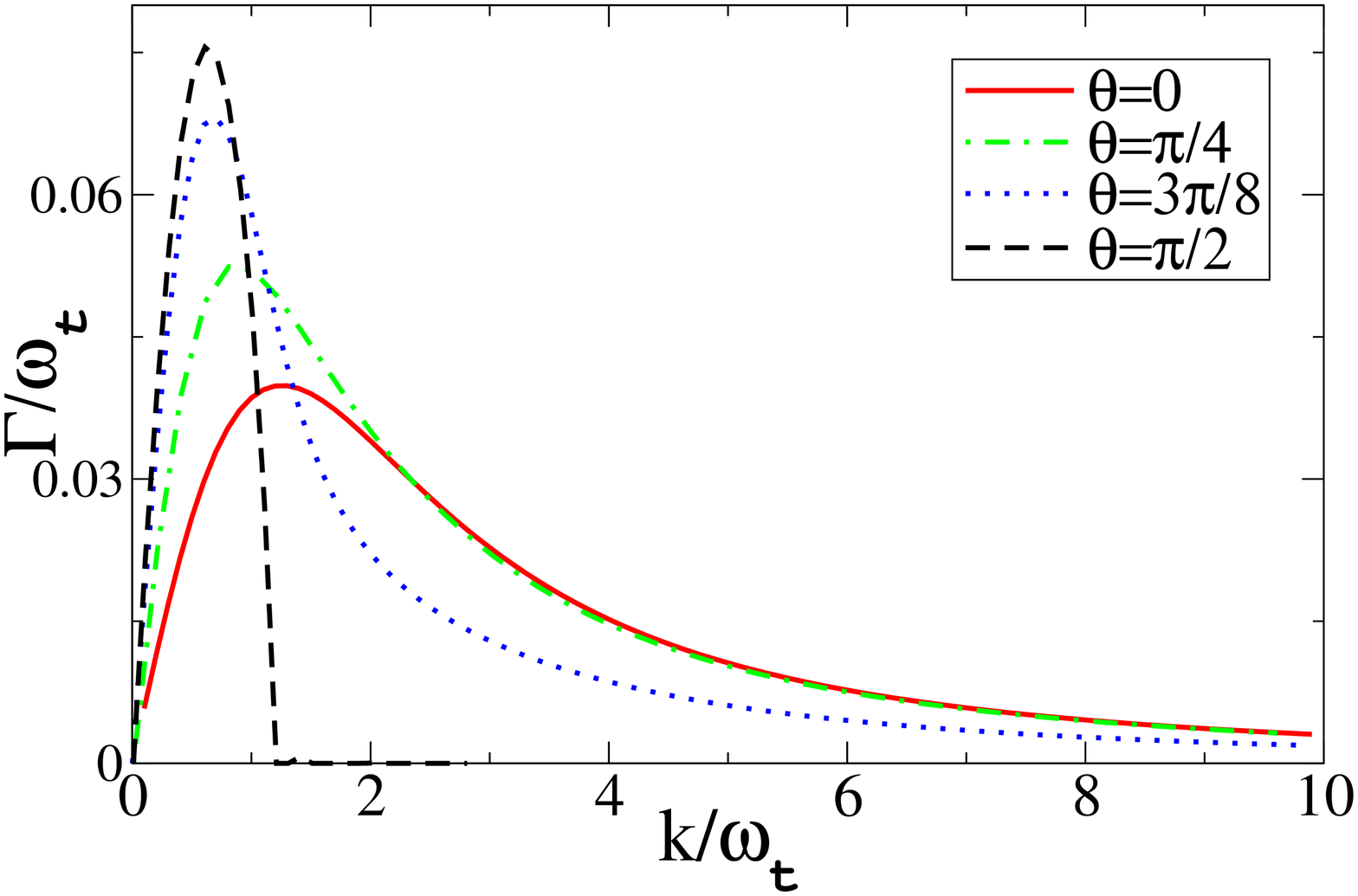}
\includegraphics[width=2.3in,angle=-0]{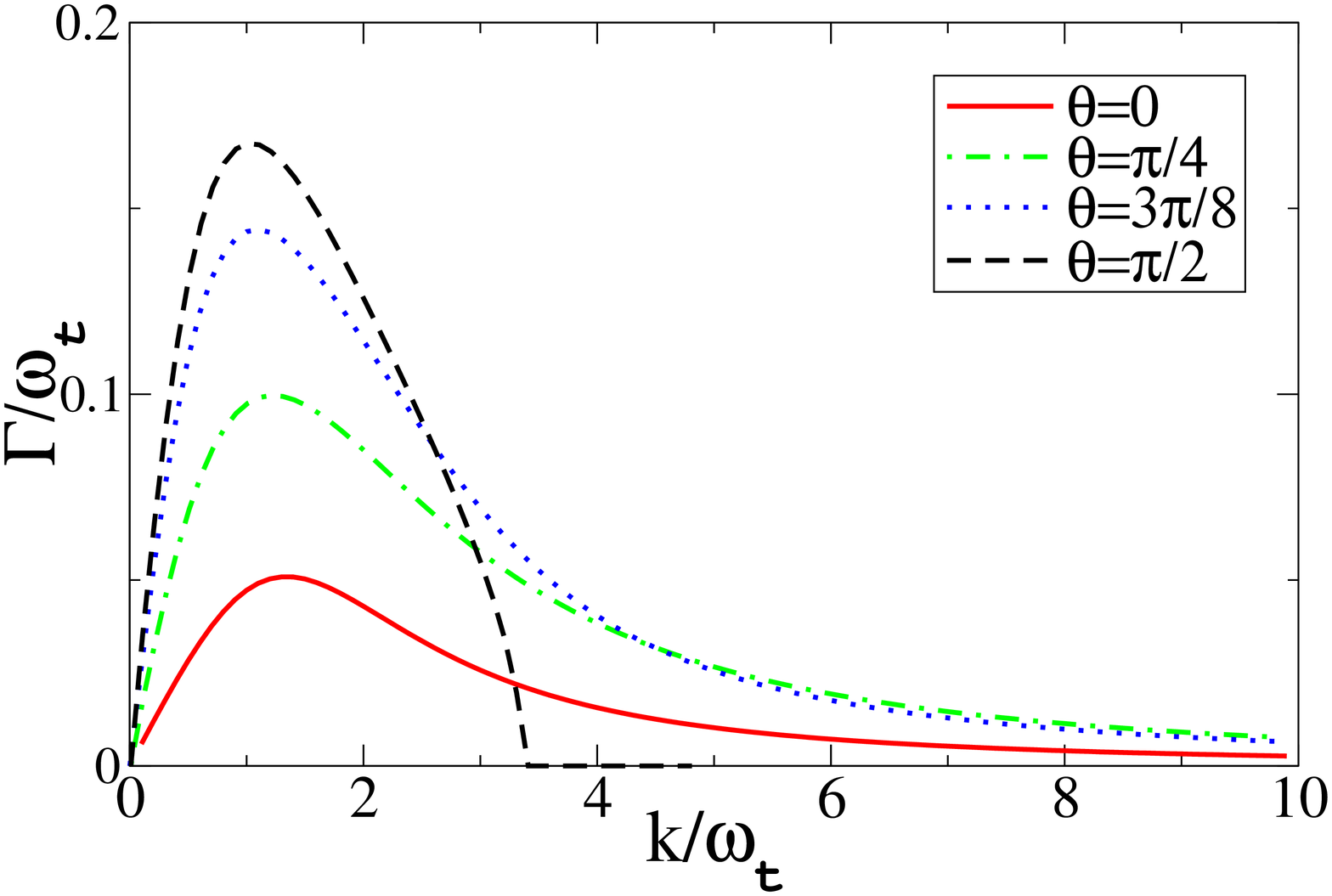}
\caption{(color online) Imaginary part of the dispersion law for the unstable mode as a function of $k$ for three different values of  the velocity of the jet ${|\bf v|}$ and four different angles between $\bf k$ and $\bf v$ for $b=0.1$. The left panel corresponds to $v=0.4$. The central panel corresponds to $v=0.6$ and the right panel to $v = 0.9$.   \vspace{1cm}} \label{obliquefig1}
\end{figure}

We have  numerically solved Eq.(\ref{eqgeneral}) and we report the   results of our  analysis  in Fig.~\ref{obliquefig1}.  Here we present the plot of the imaginary part of the dispersion law $\Gamma$ as a function of the momentum $k$ for three different values of the velocity and four different values of the angle between $\bf k$ and $\bf v$. The plots are obtained  for $b=0.1$.

For relatively small values of the velocity, the representative plot is shown on the left panel of Fig.~\ref{obliquefig1} for $v=0.4$,  there is no favored unstable direction of the momentum. Indeed the unstable modes corresponding to different angles have roughly the same maximum value of $\Gamma$.  With increasing values of the velocity of the jet, modes that are not collinear with the velocity of the jet are favored. In the  central and right panel of Fig.~\ref{obliquefig1} we show the results obtained with $v=0.6$ and $v=0.9$, showing that the largest value of the maximum of $\Gamma$ corresponds to  the mode with $\theta = \pi/2$.

\section{Fluid versus kinetic theory approaches}
\label{comparison}
In our previous paper \cite{Mannarelli:2007gi}  we discussed the appearance of
jet-induced plasma instabilities using the chromohydrodynamical approach developed in \cite{Manuel:2006hg}.
Here we compare the two formalisms, as well as the results one gets from the two approaches
for the dispersion laws of the plasma waves in different situations.

In the Vlasov approximation valid for sufficiently short time scales when the collision terms
can be neglected, the transport equations can  be re-written in a fluid-like form, as first realized for non-relativistic electromagnetic plasmas \cite{Kra73}. In Ref. \cite{Manuel:2006hg} the same strategy was followed to study the
transport equations obeyed by colored particles. Let us recall the basic inputs of such a methodology.

After defining the following two momenta moments of
the quark distribution function (antiquarks and gluons are treated similarly)
\be
\label{flow}
n^\mu (x) \equiv \int_p p^\mu Q(p,x) \;, \qquad T^{\mu \nu} (x) \equiv \int_p p^\mu p^\nu  Q (p,x) \ ,
\ee
one can obtain their dynamical evolution employing the kinetic equations.
In particular, after integrating over momenta Eq.~(\ref{transport-q}) and Eq.~(\ref{transport-q}) times
a four momentum (with $C=0$) one gets
\be
\label{fluid-eq}
D_\mu n^\mu = 0 \;, \qquad D_\mu T^{\mu \nu}
- {g \over 2}\{F_{\mu}^{\;\; \nu}, n^\mu\}= 0 \; ,
\ee
respectively.
Note that in this short time approximation there are more apparent conservation laws than in a real hydrodynamical
configuration. For example, Eqs.~(\ref{fluid-eq}) suggest that the quark particle density is a conserved quantity,
while  only baryon number (that is, the quark minus antiquark particle density) is conserved at
long time scales.

The fluid approach, originally developed to study the short time evolution of a system, might be used
at longer time scales if the effect of the collision
terms in the kinetic equations could be encoded into new terms in the fluid equations proportional to some
transport coefficients \cite{Kra73}. We will not consider such a possibility here.

In the fluid approach one considers  a specific form of the momenta moments defined above.
Assuming that the unperturbed quantities take the ideal fluid form, their small deviations are then expressed as

\be
\label{flow-lin}
n^\mu  = \bar n \, \bar u^\mu + \bar n \, \delta u^\mu
+ \delta n \,\bar u^\mu
\;,
\ee
\be
\label{en-mom-lin}
T^{\mu \nu} = (\bar \epsilon + \bar p )
\bar u^\mu \, \bar u^\nu - \bar p  \, g^{\mu \nu}
+ (\delta \epsilon + \delta p )
\bar u^\mu \, \bar u^\nu
+ (\bar\epsilon + \bar p )
(\bar u^\mu \, \delta u^\nu + \delta u^\mu \, \bar u^\nu )
- \delta p  \, g^{\mu \nu}
\;.
\ee

The relevant question now is whether the fluid approach, which essentially only studies the
dynamical evolution of two momenta moments of the distribution function, is describing
to an acceptable level of accuracy the physical situation one is interested in. 
The answer to this question depends on the situation one wants to  describe, and more specifically,
on the form of the unperturbed distribution function. In particular we are interested in seeing at which level of accuracy the fluid approach reproduces the behavior of the kinetic equations obtained in the HTL approximation. We compare in the subsequent subsections the dispersion laws obtained in the two approaches for different configurations.

\subsection{Collective modes for stable configurations}

Let us consider the physics of the beam of particles. In this case
we considered that 
the distribution
function is of the delta-like form given in Eq.~(\ref{tsunami}),  that in
the non-relativistic plasma literature is known as the ``cold beam approximation". 
For relativistic systems this tsunami-like approximation is also correct for ultrarelativistic
velocities of the jet \cite{Manuel:2006hg}.
For the unperturbed distribution function, specifying only the two momenta moments correctly characterizes
 the physics of the system. For their fluctuations 
it was found in Ref.~\cite{Manuel:2006hg} that the polarization tensor
associated to the distribution function (\ref{tsunami}) agrees exactly with the polarization
tensor one finds with the chromohydrodynamical equations by neglecting the effect of the
pressure gradients. One then concludes that in this situation the fluid methods 
describe the color polarization effects in the same way as kinetic theory.

Let us now consider the equilibrated QGP without the jet.
The distribution
functions of the particles are those of thermal equilibrium, and thus it is obvious that the unperturbed hydrodynamical variables are enough to specify the state of the system.
Can one describe the color
polarization by considering only the first and second momenta moments of $\delta Q$?
To answer this question it is instructive to compare the dispersion laws of both the transverse
and longitudinal collective modes with those obtained in the HTL approximation.
 Employing the hydrodynamical equations of Ref.~\cite{Mannarelli:2007gi} we find that
the transverse and longitudinal dispersion laws are given by
\ba
\label{tranHyd}\omega_{\rm T}^2& =& \omega_{p}^2+   k^2 \ , \\
\label{longHyd}\omega_{\rm L}^2& =& \omega_{p}^2+ (c_s^a)^2  k^2 \,,
\ea
where $c_s^a$ is defined by the relation $\delta p^a = (c_s^a)^2 \delta \epsilon^a$ \cite{Mannarelli:2007gi}. 
In a fluid approach a relation between the energy and
pressure variables is needed in order  to close the system
of equations.  In Refs. \cite{Manuel:2006hg} and \cite{Mannarelli:2007gi} it was chosen as
$c_s^a = 1/\sqrt{3}$, assuming the conformal limit. However  in a fluid description of the plasma one can consider $c_s^a$ as a free parameter. 

From the behavior of these dispersion laws,  one can infer  that the hydrodynamical approach is valid for long wavelengths, \ie for $k \ll \omega_p$. Indeed in this case the 
dispersion laws in Eqs. (\ref{tranHyd}) and (\ref{longHyd})
 approximately reproduce the  corresponding HTL dispersion laws. Also the limit $k \gg \omega_p$ seems to be approximately well
described in the fluid approach because the dispersion laws of both the fluid and the kinetic theory approach  tend  to those corresponding to
free propagators. However, for values of order $k \sim \omega_p$, we can expect some discrepancies between the fluid and kinetic approaches.

\subsection{Collective modes for the unstable configurations}

\begin{figure}[!th]
\includegraphics[width=3.5in,angle=-0]{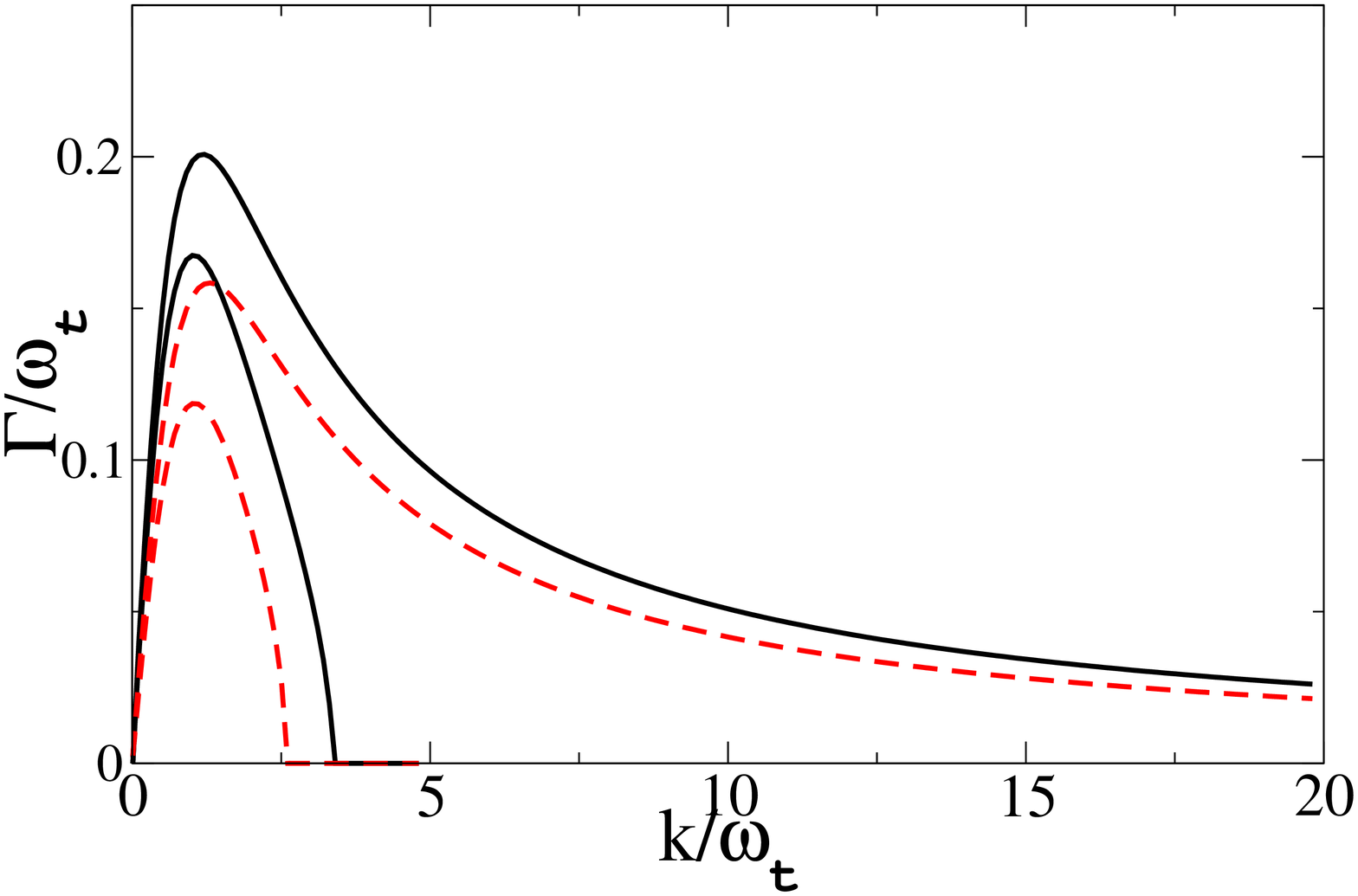}
\includegraphics[width=3.5in,angle=-0]{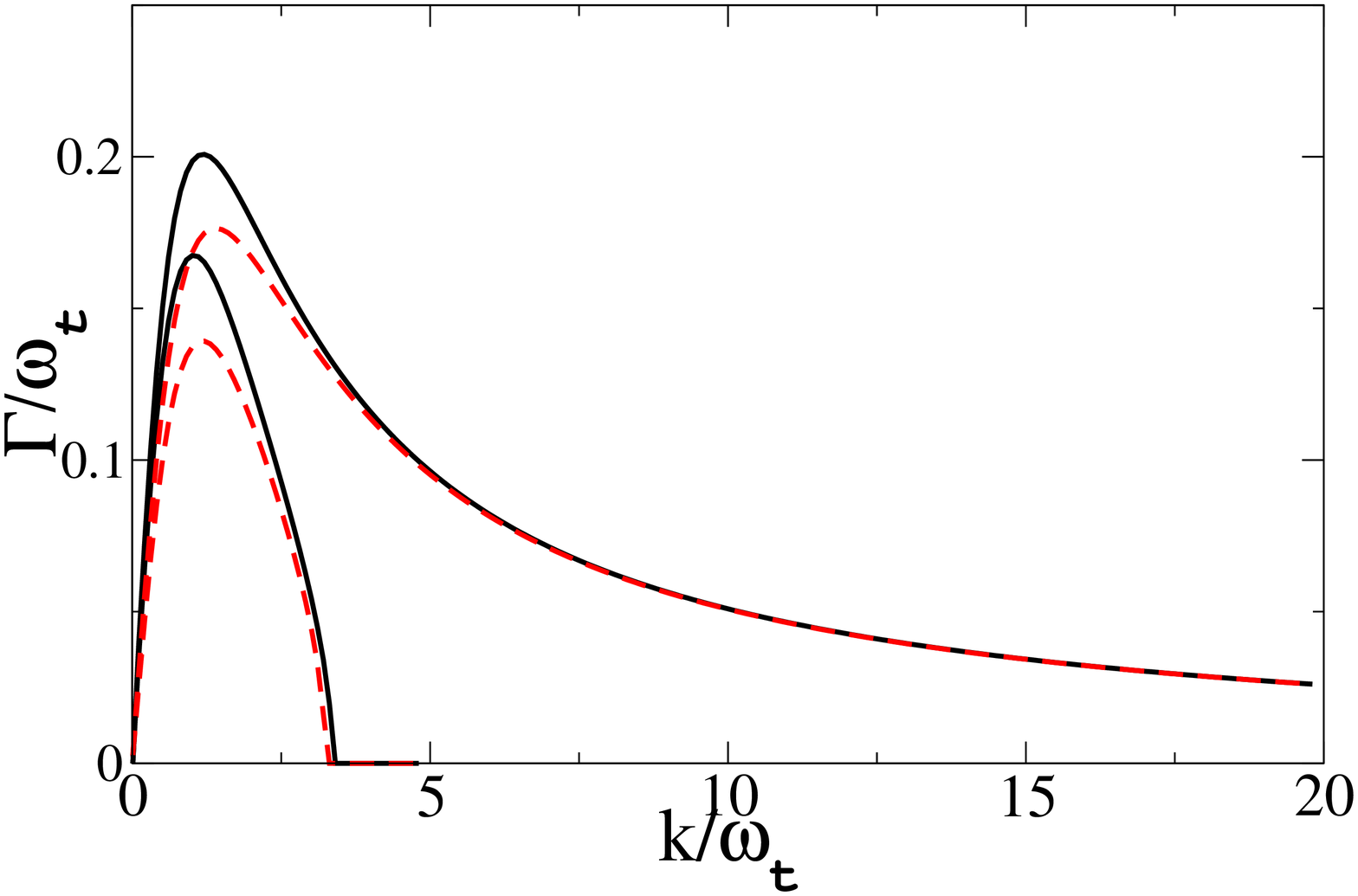}
\caption{(color online) Comparison of the imaginary part of the dispersion law of the unstable  mode for the system composed by a plasma and a jet in the case $\bf k \perp v$ as a function of the momentum of the mode at  $b=0.1$ for $v=0.9$ (lower curves) and $v=1.0$ (upper curves) in the conformal limit (left panel) and for $c_s^a= 1/2$ (right panel). Dashed (red online) lines correspond to the results obtained with the fluid approach;  full (blue online) lines  correspond to results obtained with kinetic theory. \vspace{1cm}} \label{comortofig}
\end{figure}

Let us now compare the dynamics of the equilibrated plasma traversed by a ultrarelativistic jet of particles in the 
fluid and kinetic approaches. In the two approaches unstable collective gauge modes appear.
We will first deal with transverse modes that are the dominant ones in the kinetic theory approach.  
Based on the considerations of the previous subsection, we may expect that for very short time phenomena the fluid
approach provides the correct dispersion laws  in both the situations $k \ll \omega_p$ and $k \gg \omega_p$.
We have made a comparison of the unstable growth rates obtained within the two formalisms, and this is indeed what we found. 
The results of our  analysis  for the case $\bf k \perp v$ are reported in Fig. \ref{comortofig}. 
The left panel of Fig. \ref{comortofig} corresponds to $c_s^a= 1/\sqrt{3}$, dashed lines (red online) are obtained with the fluid approach, full lines (black) are obtained  with the kinetic theory approach. We have reported the results obtained with two different values of the velocity of the jet; the upper lines correspond to $v=1$ and the lower lines correspond to   $v=0.9$. The agreement between the two approaches is quite remarkable. However it is possible to improve the agreement  employing  a different value of  $c_s^a$. In the right panel of  Fig. \ref{comortofig} we compare the results of the two methods for the  same values of the velocity of the jet, but with $c_s^a =1/2 $ \cite{footnotesound}.  The difference between the results obtained with the two methods is now of $\sim 10 \%$ at most.

Regarding the longitudinal modes, there is agreement between the fluid approach and kinetic theory
in the $k \ll \omega_p$ domain. However, beyond this region
we find that the fluid approach largely overestimates the growing rate and predicts a threshold value for the momentum.   The reason for such a disagreement is due to the fact that longitudinal modes are propagating  and the solution of Eq.~(\ref{eqparallel})
  consists of a real and an imaginary component. Therefore one cannot reproduce properly such results with one single real parameter $c_s^a$ in the regime $k \sim \omega_p$. However, it might be possible to match also this regime 
if one allows for a non-local relation between the colored pressure and energy density. This possibility is discussed in Appendix A.

There are also two qualitative differences between the results obtained within the kinetic theory method and the fluid  approach. First of all we notice that in the fluid approach the instabilities develop for velocities $v>c_s^a$ whereas in  kinetic theory the threshold value of the velocity for the development of the instability is not related to the parameter
$c_s^a$. This is clearly due to the fact that the equation of state of matter does not enter the kinetic theory picture and therefore the colored speed of sound does not play any role.

Second, in the Vlasov approximation for sufficiently small velocities, $v \lesssim 0.6 $, there is no preferred unstable direction, whereas for larger velocities   the most unstable  modes correspond to large angles
between $\bf k$ and $\bf v$. This has to be contrasted with  the results obtained using fluid equations, where one finds that for velocities $v \gtrsim  c_s^a$  the most unstable mode correspond to momenta $\bf k$ collinear with the velocity of the jet, whereas for  ultrarelativistic velocities  $v \lesssim 1  $ the unstable modes corresponding to angles $\theta \gtrsim \pi/8 $  are dominant and the most unstable mode corresponds to  $\theta \simeq \pi/4 $.

\section{Discussion}
\label{conclusion}

We have studied a system composed by an equilibrated quark-gluon plasma traversed by an energetic jet of particles
using transport theory. We have assumed that the interaction between the jet and the plasma is only mediated by mean gauge field. Moreover the plasma has been considered to be in thermal equilibrium, whereas the initial configuration
of the jet has been described employing a tsunami-like distribution function.
At very short time scales after the interaction between the jet and the plasma  sets in  we find that the effect of the jet is to destabilize the plasma  producing an exponential growth of collective gauge fields. For relatively small values of the velocity, $v \lesssim 0.4$, we do not find that any particular direction in momentum space is favored. With increasing values of $v$ the most unstable mode is characterized by having the momentum orthogonal to the direction of the velocity of the jet.  

We have also compared the results obtained in the present analysis with those obtained in \cite{Mannarelli:2007gi} where fluid equations have been employed in the description of the problem. The two formalisms describe qualitatively
the same effect: the appearance of jet-induced plasma instabilities. At very long wavelength the growth rates
obtained from the two approaches of the unstable modes match, while they differ for moderate  wavelength values.
However, the discrepancies in this region can be reduced by a proper choice of the equation of state used in the chromohydrodynamical equations.

The description of the plasma and of the jet that we have carried out is  simplistic in many ways. First of all we have assumed that the jet of particles can be described  employing a delta like distribution function. It is of course possible to employ more realistic distribution function for the particles of the jet, but this would lead to  a more complicate  numerical analysis. We have also neglected possible saturation mechanisms of the gauge instabilities. Unfortunately the  saturation time scale cannot be evaluated employing a simple linear analysis, but  requires numerical studies.   Moreover at sufficiently long times the effect of collisions should be taken into account, through the
addition of the corresponding terms in the transport equations. Indeed  if the instabilities
do not saturate due to non-linear non-Abelian effects, the collisions will probably anyway stop their growth. Further, the scatterings of the particles in the jet with those of the plasma will provide collisional and radiative jet energy loss,  which however will be relevant  at time scales longer than the mean free path time.  The consideration of all the above mentioned effects is required for a complete analysis of the possible contribution of the proposed mechanism to the jet quenching of energetic partons.

Even with the simplification employed in this paper we believe that    the effect  we discuss is relevant because it provides a new jet energy loss
mechanism, based on pure collective effects, which might be important for the description of jet quenching in heavy ion collisions.

\appendix

\section{Improved Chromohydrodynamical Formulation}
\begin{figure}[!th]
\includegraphics[width=3.5in,angle=-0]{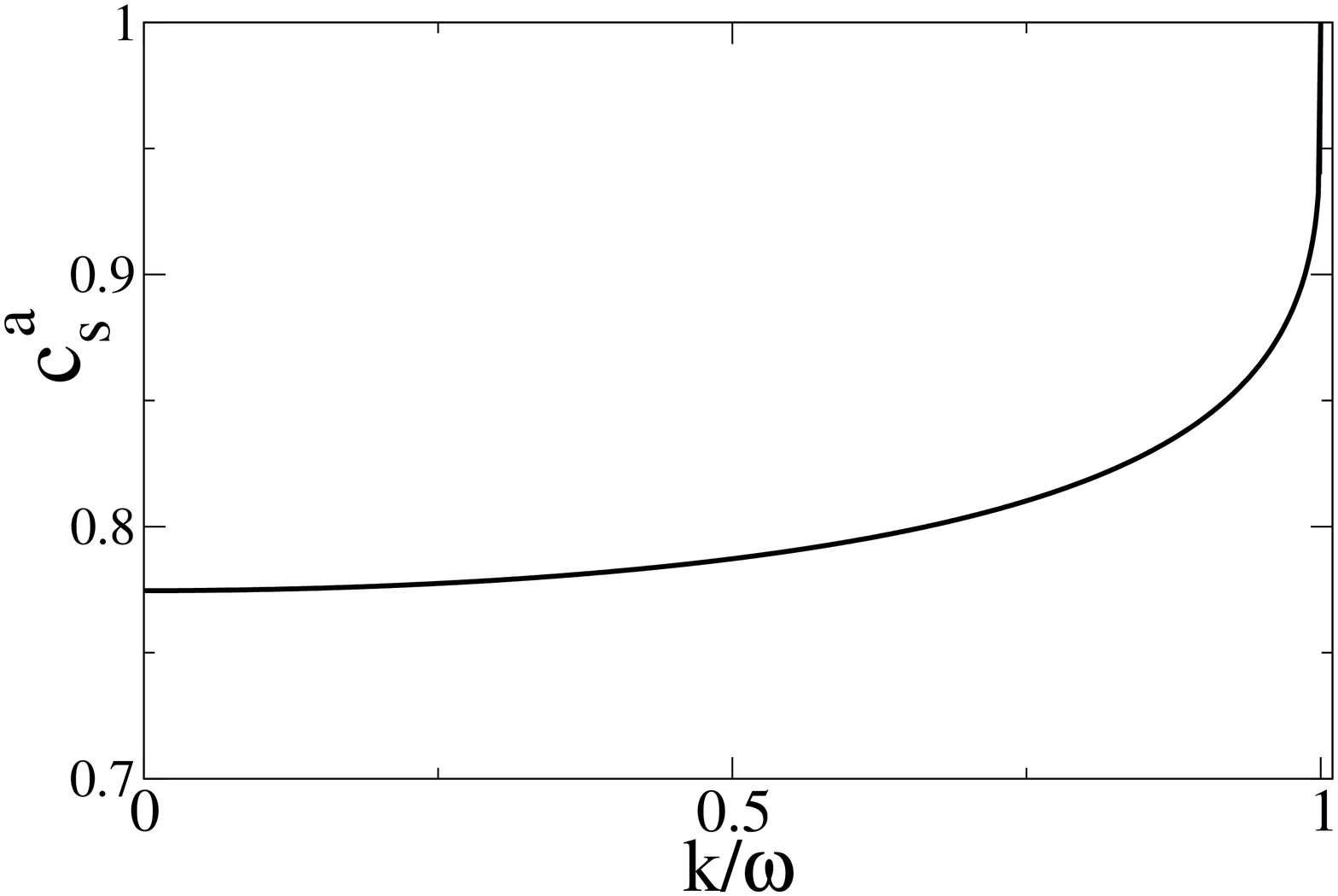}
\includegraphics[width=3.5in,angle=-0]{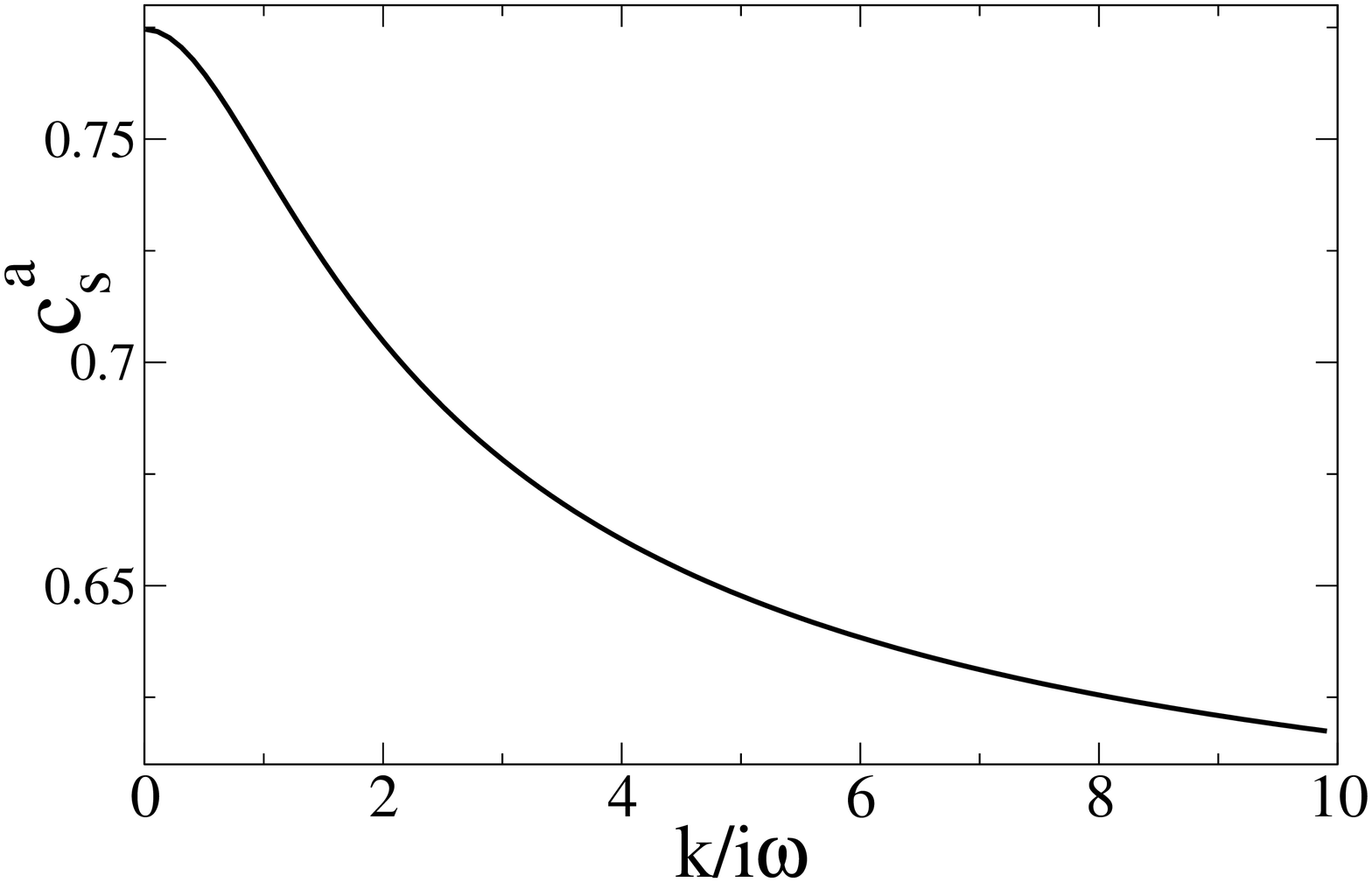}
\caption{Colored Effective speed of sound defined in Eq.~(\ref{effectivesoundeq}). In the left panel  real values of the ratio $k/\omega$ have been considered. The right panel corresponds to pure imaginary values of $k/\omega$. \vspace{1cm}}\label{effectivesound}
\end{figure}

Here  we discuss how one can improve the agreement between the results obtained with kinetic theory in the HTL approximation and the fluid approach of Ref.~\cite{Mannarelli:2007gi}.
Suppose that we want to exactly reproduce the  behavior of the longitudinal modes for the plasma without a jet in the two formalisms. In both kinetic theory and the chromohydrodynamical approach the dispersion law of the longitudinal mode is obtained by equating to zero the longitudinal component of the polarization tensor.  
In order to obtain the same dispersion law for the longitudinal modes   one can    define  an ``effective" speed of sound (we refer to the parameter which relates the colored variables $\delta p^a$ and $\delta \epsilon^a$)
in such a way that the longitudinal component of the dielectric tensor in the HTL approximation and in the fluid approach are equal.
It follows that the  ``effective" speed of sound   satisfies the equation
\be
\label{effectivesoundeq}
c_s^a(y) = \sqrt{\frac{1}{3(1+\frac{1}{2 y} \log\frac{1-y}{1+y})} + \frac{1}{y^2}}  \,,
\ee
where  $y = k/\omega$. The plot of this function is reported in Fig.\ref{effectivesound} for real values of $y$, left panel, and pure imaginary values of $y$, right panel.

For real values of $y$  the effective speed of sound is an increasing function of $y$ and  assumes  real values that are larger or equal $\sqrt{3/5}$ (and therefore larger than the conformal value) for $0 \leq y \leq 1 $. 
At $\omega=k$ this parameter reaches the largest possible value $c_s^a=1$. For values of  $\omega$ larger than $k$ the longitudinal dielectric tensor becomes imaginary and therefore it is not
possible to reproduce its behavior with a real function.  Therefore the disagreement between the results obtained with the two methods  in the description of the longitudinal modes are due to the fact that for $k \ll \omega $ the value that one should employ is not  the conformal value  $c_s^a = \sqrt{1/3} $ but $c_s^a \sim \sqrt{3/5} $, whereas for  $k > \omega $  one needs an imaginary speed of sound.

It is quite interesting that for pure imaginary values of $y$, corresponding to imaginary values of $k$ or to   imaginary values of $\omega$, the effective speed of sound is always real. It is a decreasing function of $y$ and assumes values between $\sqrt{3/5}$ and $\sqrt{1/3}$, \ie between the HTL value (in the long wavelength limit) and the conformal limit. Since transverse modes are not propagating, they correspond to pure imaginary frequencies and it is then clear that a real speed of sound suffices in approximating the behavior of these modes. However it does not explain why the value that gives a good agreement between the two approaches for the transverse mode is $c_s^a = 1/2$, that has to be determined with a different method \cite{footnotesound}.

It is worth remarking  that employing the effective speed of sound  defined above means that we are not considering the conformal limit and we  match the only free parameter of the fluid approach by means of the weak coupling theory.
This completely sets out the fluid equations.

\begin{figure}[!th]
\includegraphics[width=3.5in,angle=-0]{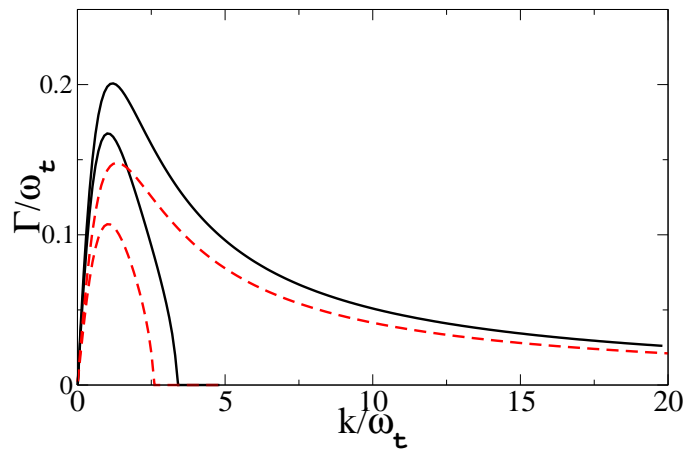}
\caption{(color online) Comparison of the imaginary part of the dispersion law of the unstable  mode for the system composed by a plasma and a jet in the case $\bf k \perp v$ as a function of the momentum of the mode at  $b=0.1$ for $v=0.9$ (lower curves) and $v=1.0$ (upper curves) employing the effective speed of sound defined in Eq.~(\ref{effectivesoundeq}). Dashed (red online) lines correspond to results obtained with the fluid approach;  full (black) lines  correspond to results obtained with the kinetic theory approach. \vspace{1cm}} \label{comparaeffe}
\end{figure}

We can  now compute the dispersion laws of the  longitudinal and transverse modes for the system composed by the plasma traversed by  the jet employing the  effective speed of sound defined in  Eq. (\ref{effectivesoundeq}). Since the effective speed of sound is defined in such a way that the longitudinal component of the polarization tensor of the plasma in the HTL and fluid approximations are the same, it follows from Eq.(\ref{eqparallel}) that the longitudinal modes for the system composed by the plasma and the jet are exactly the same. Regarding the transverse modes,   the results of the fluid approach are compared with those of the HTL approximation in Fig \ref{comparaeffe}.  

The agreement is quite satisfactory, but is not as good as in Fig. \ref{comortofig}. The reason being that the ``preferred" value of the speed of sound for the transverse mode is $c_s^a= 1/2$, whereas we are employing the ``effective" speed of sound reported in Eq.~(\ref{effectivesoundeq}). As can be seen from the right hand side of Fig.~\ref{effectivesound} the values of such a function are between $\sqrt{3/5}$ and  $\sqrt{1/3}$ and are therefore larger than the value $c_s^a= 1/2$.

\begin{acknowledgments}
This work has been supported by the Ministerio de Educaci\'on y Ciencia (MEC) under grant AYA 2005-08013-C03-02.

\end{acknowledgments}


\end{document}